# Electric-field-induced parametric excitation of exchange magnons in a CoFeB/MgO junction

Angshuman Deka[1#], Bivas Rana[2,3], Ryo Anami[1], Katsuya Miura[4], Hiromasa Takahashi[4], YoshiChika Otani[2,5], and Yasuhiro Fukuma[1,2,6*]

[1] *Department of Physics and Information Technology, Kyushu Institute of Technology, 680-4 Kawazu, Iizuka 820-8502, Japan*

[2] *Center for Emergent Matter Science, RIKEN, 2-1 Hirosawa, Wako 351-0198, Japan*

[3] *Institute of Spintronics and Quantum Information, Faculty of Physics, Adam Mickiewicz University in Poznań, Uniwersytetu Poznanskiego 2, Poznań 61-614, Poland*

[4] *Research and Development Group, Hitachi Ltd., 1-280 Higashi-koigakubo, Kokubunji 185-8601, Japan*

[5] *Institute for Solid State Physics, University of Tokyo, Kashiwa, Chiba 277-8581, Japan*

[6]*Research Center for Neuromorphic AI hardware, Kyushu Institute of Technology, Kitakyushu 808-0196, Japan*

[#]Present address: *Birck Nanotechnology Center, School of Electrical and Computer Engineering, Purdue University, West Lafayette, Indiana 47907, USA.*

[*]Correspondence should be addressed to Y.F. (fukuma@cse.kyutech.ac.jp)

**Abstract**

Electric-field controlled magnetization dynamics is an important integrant in low power spintronic devices. In this letter, we demonstrate electric-field induced parametric excitation for CoFeB/MgO junctions by using interfacial in-plane magnetic anisotropy. When the in-plane magnetic anisotropy and the external magnetic field are parallel to each other, magnons are efficiently excited by using electric-field induced parametric excitation. Its wavelength and wavenumber are tuned by changing input power and frequency of the applied voltage. A generalized phenomenological model is developed to explain the underlying role of the electric-field torque. Electrical excitation with no Joule heating offers a good opportunity for developing magnonic devices and exploring various nonlinear dynamics in magnetic systems.

**Keywords:** spintronics, voltage controlled magnetic anisotropy, electric-field torque, ferromagnetic resonance, parametric excitation, magnons.

Nonlinear dynamics in a physical system has attracted considerable attention because of their fundamental interest and potential application for reservoir computing [1-3]. In magnetic materials, the interaction among spins is intrinsically nonlinear and several phenomena such as harmonic oscillation and auto-oscillation can be observed for their excited state [4,5]. The collective excitation of spins gives rise to propagating spinwave modes, also quantum mechanically known as magnons, which carry angular momentum. Therefore, such nonlinear dynamics is also useful for generating spin current [6-8]. Magnetic insulator such as $Y_3Fe_5O_{12}$ (YIG) is considered an excellent material for studying the nonlinear dynamics due to small magnetic damping. However, unlike the high quality garnets grown on $Gd_3Ga_5O_{12}$ substrates, existing CMOS technology is based on metallic thin films grown on Si substrates. An ability to excite and control of nonlinear dynamics in metallic ferromagnets such as NiFe and CoFeB could be a promising avenue towards realizing spintronic devices that are compatible with the CMOS processes.

Reports on the nonlinear parametric excitation in metallic ferromagnets are less compared to insulating magnets like YIG because of higher linewidths of resonant spectrum [9-11], which in turn increases the threshold power required for exciting nonlinear dynamics. Recent discoveries such as spin–transfer torque [12,13], spin-orbit torque [14] and electric-field torque [15] in the field of spintronics provide a new strategy for studying nonlinear dynamics in magnetic systems. In particular, the electric-field torque, originating from voltage controlled magnetic anisotropy (VCMA) at ferromagnetic metal/oxide interfaces, allows us to control the magnetization without Joule heating [15-18]. The parametric excitation using VCMA was theoretically predicted [19], and then was experimentally realized by using magnetic tunnel junctions with a MgO tunneling layer [20]. VCMA originates from the selective electron-hole doping in the $3d$ orbitals of the interfacial ferromagnetic atoms in the presence of electric field [21,22]. Therefore, its electric-field torque on the magnetization arises due to a modulation of the interfacial perpendicular magnetic anisotropy (PMA) [23-28]. The torque is proportional to $(dH_p/dV)V \sin\theta \cos\theta$: modulation of PMA field $H_p$ per unit voltage $V$ i.e. $(dH_p/dV)V$ and orientation of magnetization i.e. $\theta$ is the elevation angle of magnetization from the film plane. Note that the torque on the magnetization is zero for in-plane ($\theta = 0$ degree) and perpendicular ($\theta = 90$ degree) magnetized ferromagnetic thin films. In order to overcome this limitation, the interfacial magnetic anisotropy of CoFeB/MgO junctions can be tuned by using magnetic annealing [29], which allows us to control the symmetry of the electric-field torque via VCMA (see Eq. (2)). In this study, we demonstrate electric-field-induced magnetization dynamics for an in-plane magnetized CoFeB layer by using VCMA of in-plane magnetic anisotropy (IMA) induced by magnetic annealing. The parallel pumping configuration, in which the microwave driving field generated by voltage-controlled IMA is applied parallel to the external field, has lower critical fields of nonlinear magnetization dynamics which can then excite various magnons. The magnons are detected using spin pumping and inverse spin Hall voltage measurements. A phenomenological model is developed to confirm the role of the electric-field torque via VCMA of IMA.

Multilayer stacks of Ta(5)/Ru(10)/Ta(5)/Co$_{20}$Fe$_{60}$B$_{20}$(1.8)/MgO(2)/Al$_2$O$_3$(10) were sputtered at room temperature onto thermally oxidized Si (100) substrates. The multilayer stack was patterned into rectangles of length (*l*) 250 μm and width (*w*) 20 μm by using photolithography and Ar$^+$ ion milling. Au electrodes were fabricated by lift-off processes. Post device fabrication, the samples were annealed along the shorter axis of the rectangle (x axis, $\theta = \phi = 0$ degree) to induce an in-plane magnetic anisotropy. The magnetization dynamics is excited by applying AC voltage which modulate $H_p$ and $H_k$ of the CoFeB/MgO junction, as shown schematically in Fig. 1(a). Uniform ferromagnetic resonance (FMR) and magnons can generate spin current toward an adjacent Ta layer via spin pumping and the injected spin current can be detected using inverse spin Hall effect (ISHE). The ISHE voltage $V_{ISHE}$ was detected by a standard lock-in detection technique. The amplitude of the input microwave is modulated at a reference frequency of 79 Hz. The DC rectified voltage is detected across the two ends of the Ta strip (bottom electrodes marked on the upper side in Fig. 1(b)) using a lock-in amplifier. We obtained $\mu_0 H_p$ = 1405 mT and $\mu_0 H_k$ = 2.5 mT for PMA and IMA, respectively, from $\phi$ dependent $V_{ISHE}$ measurements [30]. All measurements are performed at room temperature.

Figure 2 (a) shows a typical $V_{ISHE}$ spectrum as a function of input power $P_{in}$ at frequency $f$ = 3 GHz when $H_{ex}$ is applied along $\theta = \phi = 0$ degree. For $P_{in}$ < 0.08 W, we observe a single peak around $\mu_0 H_{ex}$ = 63 mT ($H_{res}^1$) in the $V_{ISHE}$ spectrum. As $P_{in}$ increases above 0.08 W, we observe the emergence of a second peak around $\mu_0 H_{ex}$ = 23 mT ($H_{res}^2$). Figure 2(b) shows the $P_{in}$ dependence of $H_{res}^1$. Above $P_{in}$ ~ 0.08 W, $H_{res}^1$ starts to shift. A second peak also appears at $H_{res}^2$ above $P_{in}$ ~ 0.08 W, indicating onset of parametric resonance excitation. The $V_{ISHE}$ spectrum at $P_{in}$ = 0.3 W with different applied microwave frequencies is shown in Fig. 2(c). $H_{res}^1$ and $H_{res}^2$ increase with increasing frequency and are well fitted to the Kittel formula, $f = \gamma\sqrt{H_{res}^1(H_{res}^1 + M_{eff})}$ and $\frac{f}{2} = \gamma\sqrt{H_{res}^2(H_{res}^2 + M_{eff})}$, respectively, with $\gamma$ the gyromagnetic ratio whose value is estimated to be 0.0298 GHz.mT$^{-1}$ [29] and $\mu_0$ the permeability of free space. The effective demagnetizing field $M_{eff}$ (= $M_s - H_p$) is estimated to be 95 mT. The rectified voltage at $H_{res}^1$ and $H_{res}^2$ corresponds to the uniform FMR mode and parametric magnon mode, respectively.

The threshold power of the parametric excitation as a function of the external magnetic field was measured [30], as shown in Fig. 3 (a). We can see the lowest threshold power at $H_{ex}$ ~ $H_{res}^2$ and a well-known butterfly curve in the nonlinear dynamics reported for YIG and NiFe [31- 33]. In the linear excitation regime, the cone angle of the magnetization precession is small enough to consider that $M_{eff}$ is constant and therefore no change of $H_{res}^1$ is observed in the $P_{in}$ dependence, as can be seen in Fig. 2(b). However, in the nonlinear excitation regime, a large angle of precession causes a change of $M_{eff}$, evident from $H_{res}^1$ that changes with $P_{in}$ [34, 35]. The magnetization component parallel to the effective field i.e., $m_x$, oscillates ($\delta m_x$) at a frequency twice that of the transverse components [36]. This can directly couple with a modulation of $H_k$ via VCMA, so-called parallel pumping, which shows a minimum of the threshold power at $H_{ex} \sim H_{res}^2$. As $P_{in}$ increases, various magnon modes are excited for $H_{ex} < H_{res}^2$. We can calculate the corresponding wavelength λ as a function of $H_{ex}$ using the relation [37]:

$$\lambda = \left[ \frac{M_{eff}}{2\gamma A} \left( \sqrt{\left(\frac{f}{2}\right)^2 + \left(\frac{\gamma M_{eff}}{2}\right)^2} - \gamma \left( H_{ex} + \frac{M_{eff}}{2} \right) \right) \right]^{-0.5}. \quad (1)$$

Figure 3 (b) shows the $H_{ex}$ vs λ plot for the sample in this study. It can be seen that the obtained value of λ in the low $H_{ex}$ region are 2 orders smaller than the lateral excitation length scales, well into the exchange-interaction regime for magnons which becomes predominant at λ < 1μm. Its wavelength and wavenumber can be tuned by changing input power and frequency of the applied voltage. The drastic increase of the threshold power for $H_{ex} > H_{res}^2$ is because no modes exist in this region.

Figures 4(a) and 4(b) show the $P_{in}$ dependence of the magnitude of $V_{ISHE}$ at $\mu_0 H_{ex}$ = 31, 11.5, 6 and 3 mT for $f$ = 2 GHz, and at $\mu_0 H_{ex}$ = 63.1, 20.5, 10 and 5 mT for $f$ = 3 GHz, respectively. In low $P_{in}$ region, $V_{ISHE}$ of the FMR mode ($\mu_0 H_{ex}$ = 31 and 63.1 mT) increases linearly with increasing $P_{in}$ and no broadening of the $V_{ISHE}$ peak was observed, implying the magnetization dynamics follows the Landau-Lifshitz-Gilbert equation with a constant damping term in the linear exaction regime. As $P_{in}$ increases further, an additional peak due to the excitation of parametric resonance can be observed around $\mu_0 H_{ex}$ = 11.5 and 20.5 mT for $f$ = 2 and 3 GHz, respectively. Note that $V_{ISHE}$ of the parametric excitation mode increases linearly with increasing $P_{in}$ in the low $P_{in}$ region, as can be seen in Fig. 4. VCMA of $H_k$ which is parallel to the external magnetic field can directly couple with a quasi-uniform mode at 2*f*, so-called parallel pumping, and therefore the $V_{ISHE}$ increases linearly with $P_{in}$ which can be attributed to a monotonic increase in the amplitude of $\delta m_x$. In high $P_{in}$ region, such a linear behavior for the parametric excitation mode is suppressed and then non-zero $V_{ISHE}$ at low magnetic fields ($\mu_0 H_{ex}$ = 6 and 3 mT in Fig. 4(a), 10 and 5 mT in Fig. 4(b)) are observed. This behavior can be understood in terms of energy transfer from the parametric mode to higher-order magnon modes, in which the threshold power of the instabilities is determined by competition between the mode-coupling efficiency and the magnon linewidth [38]. The coupling efficiency scales with the square of its amplitude to the parametric excitation mode [31], and therefore $V_{ISHE}$ increases as a function of the square of $P_{in}$. As $P_{in}$ increases further, there is a deviation from linear dependence, which may be due to scaling of the voltages as a function of the square root of the power because of the temperature independence of parametric instabilities [20]. We note that in our samples, such a deviation from a linear dependence at high input power may also be due to Joule heating and magnon-magnon scattering which increases the magnon linewidth.

In order to understand VCMA contribution to the magnetization dynamics in the samples, we develop a phenomenological model of $V_{ISHE}$. The torque arising from VCMA of both $H_p$ and $H_k$ can be estimated as below [30]:

$$\tau_{VCMA} \propto \begin{bmatrix} -\frac{\partial H_p}{\partial V} V_{rf} \sin\phi \sin\theta \cos\theta \\ (\frac{\partial H_p}{\partial V} - \frac{\partial H_k}{\partial V}) V_{rf} \cos\phi \sin\theta \cos\theta \\ \frac{\partial H_k}{\partial V} V_{rf} \sin\phi \cos\phi \cos^2\theta \end{bmatrix}, \qquad (2)$$

where, $\partial H_p/\partial V$ and $\partial H_k/\partial V$ represents the voltage dependence of anisotropy fields $H_p$ and $H_k$, respectively, and $V_{rf}$ is the amplitude of AC voltage applied at the CoFeB/MgO junction. From a theoretical model, rectified voltage of $V_{ISHE}$ is given by $V_{ISHE} = A' \theta_c^2 \cos\phi \cos\theta$ [39-41]. Here, $\theta_c$ is the cone angle of the magnetization precession and $A' = -\frac{\theta_{SH} e \zeta f \lambda_{sd} L_{NM} g^{\uparrow\downarrow}}{\sigma_{NM} t_{NM}} tanh\left(\frac{t_{NM}}{2\lambda_{sd}}\right)$,

respectively, where $e$ and $g^{\uparrow\downarrow}$ are the charge of an electron and spin mixing conductance of the Ta/CoFeB interface, respectively. Here, $\varsigma$, $\theta_{SH}$, $\lambda_{sd}$, $\sigma_{NM}$, $L_{NM}$ and $t_{NM}$ are the correction factor from arising from elliptical precession trajectories, spin Hall angle, spin diffusion length, conductivity, length and thickness of the Ta layer. In our analysis of the angular dependence of the uniform FMR mode, since our $H_p$ is almost equal to the demagnetizing field, we can approximate the $\varsigma \sim 1$ [36]. In the linear excitation regime, $\theta_c = \eta\,\tau$, we can obtain the following expression of $V_{ISHE}$ for the electric-field torques:

$$V_{ISHE,VCMA} = A V_{rf}^2 \left(\left(\frac{\partial H_p}{\partial V}\sin\phi\sin\theta\right)^2 + \left(\left(\frac{\partial H_p}{\partial V} - \frac{\partial H_k}{\partial V}\right)\cos\phi\sin\theta\right)^2 + \left(\frac{\partial H_k}{\partial V}\sin\phi\cos\phi\cos\theta\right)^2\right)\cos\phi\cos^3\theta. \quad (3)$$

where $A = A'\eta^2$ and $\eta$ is a constant of proportionality correlating $\theta_c$ and $\tau$. The application of a microwave voltage in the sample gives rise to an induced current, which causes $h_{rf}$ [23]. Therefore, we should consider Oersted-field torques caused by $h_{rf,x}$, $h_{rf,y}$ and $h_{rf,z}$ and performed finite element method (FEM) analysis using COMSOL Multiphysics simulator [30]. It reveals the presence of x- and y-axial Oersted field contributions arising from an induced current, while the z-axial Oersted field is zero. Further, for the sample used in this study, which have $l:w$ = 12:1, the Oersted field along the shorter in-plane dimension is very small compared to those oriented along the longer one, i.e. $h_{rf,x} \ll h_{rf,y}$ [30]. As a result, the in-plane angle dependence ($\theta = 0$ degree) of $V_{ISHE}$ can be obtained as follows:

$$V_{ISHE,\phi} = A\left(4\left(\frac{\partial H_k}{\partial V}V_{rf}\right)^2\sin^2\phi + h_{rf,y}^2\right)\cos^3\phi. \quad (4)$$

Figure 5 (a) shows $V_{ISHE}$ as a function of $\phi$ for the FMR mode. While the Oersted-field torque arising from $h_{rf,y}$ gives us a maxima at $\phi = 0$ degree, the electric-field torque via VCMA of $H_k$ gives us a maxima at an angle intermediate to $\phi = 0$ and 90 degrees. Therefore, an increase of $V_{ISHE}$ at $\phi = 40$ degree in Fig. 5(a) suggests that the electric-field torque gives rise to the FMR mode. The experimental data are well fitted using Eq. (4) with the values $A\left(\frac{\partial H_k}{\partial V}V_{rf}\right)^2$ = 2.87 and $Ah_{rf,y}^2 = 0.88$. The contribution of $h_{rf,y}$ can be understood to arise from an impedance mismatch problem between the sample and the 50 Ω signal line. As the frequency increases, the VCMA efficiency decreases and the Oersted-field contribution increases because the sample shows a capacitive nature [30]. This was confirmed by performing the angular dependences at different frequencies; we obtained values $A\left(\frac{\partial H_k}{\partial V}V_{rf}\right)^2 = 0.90$ and $Ah_{rf,y}^2 = 0.65$ for $f = 4$ GHz [30]. In this study, at $P_{in} = 0.001$ W, we confirm that the electric-field torque is higher than the Oersted-field torque up to applied microwave frequency ~ 4 GHz.

In addition to the uniform FMR mode, we studied the torque contribution of the parametric excitation. Figure 5 (b) shows $V_{ISHE}$ as a function of $\phi$ at $\mu_0 H_{ex} = 11.5$ mT for $f = 2$ GHz. The amplitude can be fitted using a $\cos^3\phi$ function, which is different from the FMR mode. This behavior is well understood by considering parallel pumping mechanism. VCMA of $H_k$ can directly couple with the quasi-uniform mode with $2f$ and thus its torque is obtained to be a function of $\cos\phi$. In a similar way to the discussion above, assuming precession angle $\theta_c$ is proportional to its torque, $V_{ISHE}$ is proportional to $\theta_c^2$ and thus $\cos^2\phi$. The product of this with a detection efficiency of ISHE which is proportional to $\cos\phi$, gives us a total of $\cos^3\phi$ that is used for fitting to the experimental data. A similar angular dependence is obtained for other frequencies. For the parametric excitation, two mechanisms can be considered: one is the parallel pumping [31] and the other is the first order Suhl instability via the FMR mode [42]. As mentioned above, the parallel pumping field can be directly coupled with the magnon mode and therefore the critical field of the parametric excitation is much lower than that of the first order process. In fact, no magnon with $k \neq 0$ is excited by using the Oersted-field torque of $h_{rf,z}$ in an control sample, that can give rise to only perpendicular pumping mechanism [30]. To summarize, in Fig. 2 – 4 when $H_{ex}$ is along x-axis, although the uniform FMR mode is excited by Oersted-field $h_{rf,y}$, the excitation of exchange magnons via parametric resonance can only be explained using a parallel pumping mechanism i.e. VCMA of $H_k$.

In conclusion, we have studied the excitation of magnetization dynamics in CoFeB/MgO junctions by using electric-field torque via voltage controlled magnetic anisotropy. The interfacial in-plane and out-of-plane magnetic anisotropies allows us to excite not only uniform ferromagnetic resonance mode but also magnon mode with the nanoscale wavelength for the in-plane magnetized CoFeB layer. More importantly, through the VCMA of in-plane magnetic anisotropy, parallel pumping allows us to reach critical fields at lower powers for the nonlinear excitation and couple effectively with various magnon modes, which allow us to tune its wavelength and wavenumber. Magnetoelectric effects in ferromagnetic metal-oxide junctions are reported through other mechanisms such as reversible oxidation [43] and piezoelectricity [44], in addition to the voltage controlled magnetic anisotropy. The advantage in this technique lies in its compatibility with existing CMOS technologies that operate at $f > 10^9$ Hz regime, whereas the former effects operate only in the few $10^3$ Hz range. Therefore, the electric-field torque in this study is useful for developing magnonic logic devices [45].

We would like to thank K. Kondou for helpful discussions. This work was supported by Grants-in-Aid (No. 26103002, 16K18079, 18H01862) from JSPS and Materials Science Foundation, Hitachi Metals. B.R. acknowledges RIKEN Incentive Research Project Grant No. FY2019.

**Figure legends:**

**Fig. 1. (a)** Schematic showing the orientation of the perpendicular magnetic anisotropy field $H_p$ and in-plane magnetic anisotorpy field $H_k$. Magnetization $M$ is taken parallel to magnetic field $H_{ex}$ applied at an elevation angle $\theta$ and azimuthal angle $\phi$. A microwave voltage is applied at the CoFeB/MgO junction using an input power $P_{in}$. Spin current $J_s$ is pumped and then is converted into charge current $J_c$ due to inverse spin Hall effect of the Ta layer. **(b)** Optical image of the sample used in this study. The top electrode on MgO is marked as T and the bottom electrodes of the Ta layer are marked as B. The rectified voltage from $J_c$ i.e. $V_{ISHE}$ is detected across the Ta/Ru/Ta underlayer.

**Fig. 2. (a)** Inverse spin Hall voltage $V_{ISHE}$ as a funciton of external magnetic field $H_{ex}$ for different input power $P_{in}$ at a frequency $f = 3$ GHz when $H_{ex}$ is applied along $\theta = \phi = 0$ degree. **(b)** Resoance field $H_{res}^1$ of the ferromagnetic resonace mode as a function of $P_{in}$ at $f = 3$ GHz. **(c)** $V_{ISHE}$ spectra at high power $P_{in} = 0.3$ W and $f = 2 - 5$ GHz. **(d)** $H_{res}^1$ and $H_{res}^2$ as a function of $f$ with the fits to the Kittel equation.

**Fig. 3. (a)** Threshold power of parametric excitation modes as a function of $H_{ex}$ for $f = 2 - 4$ GHz. **(b)** Corresponding magnon wavelenths calculated using Eq. (1) as a function of $H_{ex}$ for $f = 2 - 4$ GHz.

**Fig. 4.** Amplitudes of $V_{ISHE}$ at different $H_{ex}$ as a function of input power $P_{in}$ at different frequencies **(a)** 2 GHz and **(b)** 3 GHz. Grey and red lines are linear fits to $P_{in}$ and blue and green lines are fits to a square of $P_{in}$.

**Fig. 5. (a)** Dependence of $V_{ISHE}$ amplitudes on azimuthal angle $\phi$ at $\mu_0 H_{ex}$ = 36 mT for input power of $P_{in}$ = 0.001 W and $f$ = 2GHz. Line is fit to Eq. (4). **(b)** Dependence of $V_{ISHE}$ amplitudes on azimuthal angle $\phi$ at $\mu_0 H_{ex}$ = 11.5 mT for input power of $P_{in}$ = 0.280 W and $f$ = 2GHz. Line is fit to $\cos^3\phi$.

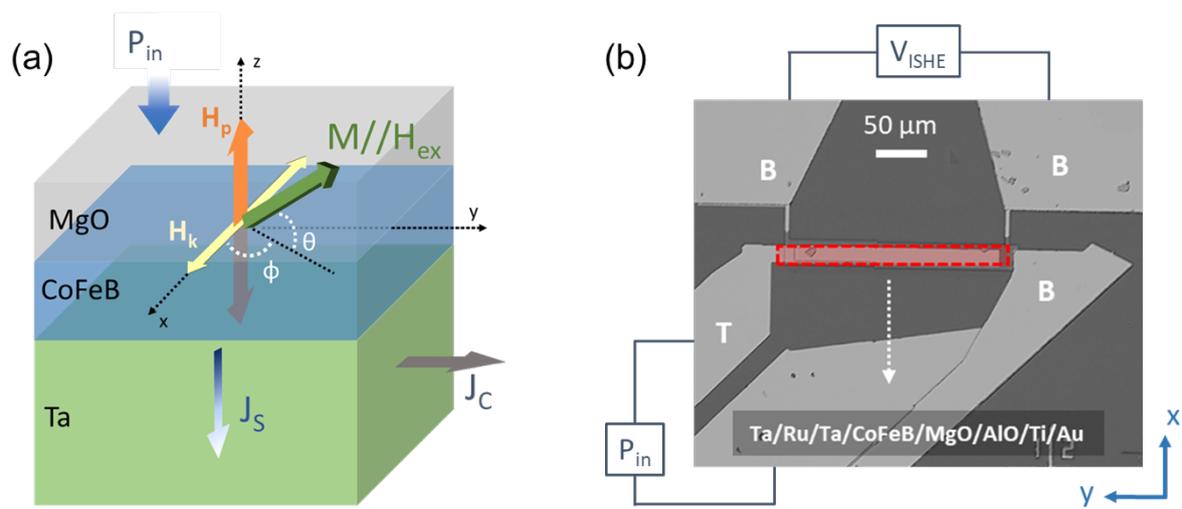

Fig. 1, A. Deka *et al.*

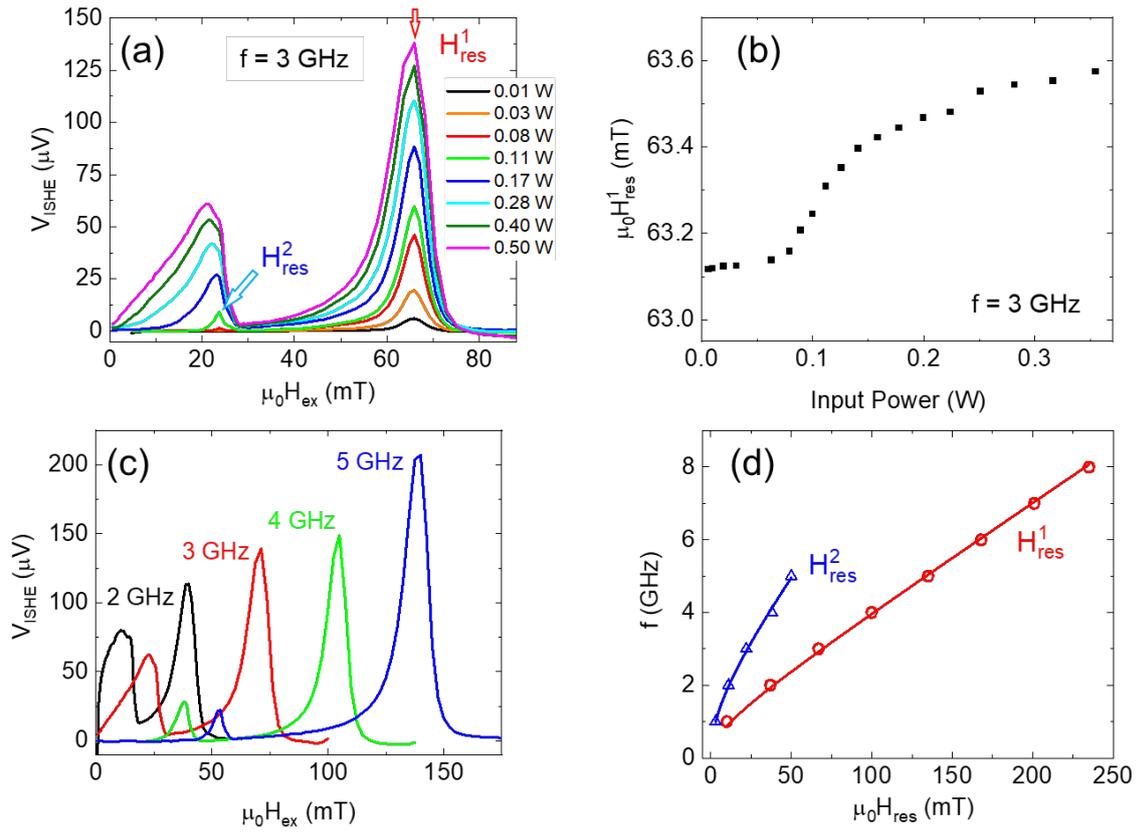

Fig. 2, A. Deka *et al.*

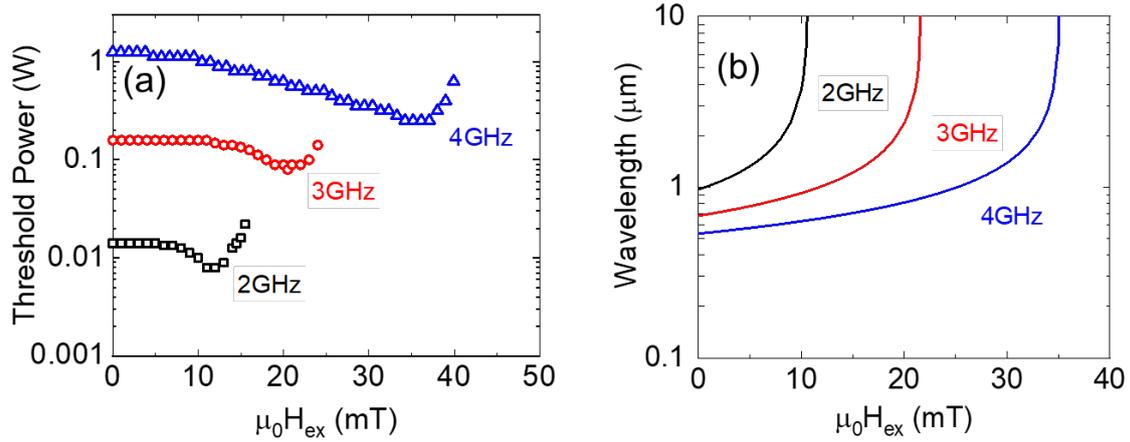

Fig. 3, A. Deka *et al.*

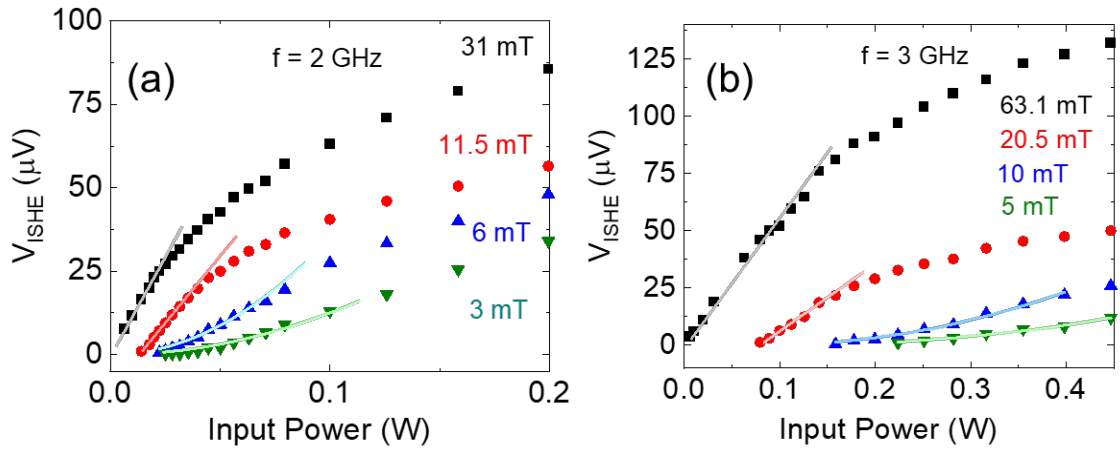

Fig. 4, A. Deka *et al.*

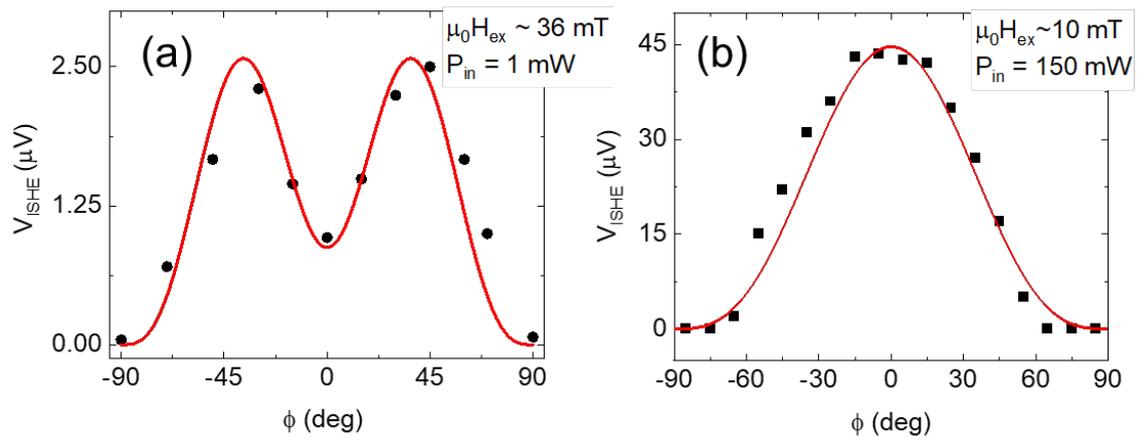

Fig. 5, A. Deka *et al.*

# Supplemental Materials

# Electric-field-induced parametric excitation of exchange magnons in a CoFeB/MgO junction

Angshuman Deka[1#], Bivas Rana[2,3], Ryo Anami[1], Katsuya Miura[4], Hiromasa Takahashi[4], YoshiChika Otani[2,5], and Yasuhiro Fukuma[1,2,6*]

[1] *Department of Physics and Information Technology, Kyushu Institute of Technology, 680-4 Kawazu, Iizuka 820-8502, Japan*
[2] *Center for Emergent Matter Science, RIKEN, 2-1 Hirosawa, Wako 351-0198, Japan*
[3] *Institute of Spintronics and Quantum Information, Faculty of Physics, Adam Mickiewicz University in Poznań, Uniwersytetu Poznanskiego 2, Poznań 61-614, Poland*
[4] *Research and Development Group, Hitachi Ltd., 1-280 Higashi-koigakubo, Kokubunji 185-8601, Japan*
[5] *Institute for Solid State Physics, University of Tokyo, Kashiwa, Chiba 277-8581, Japan*
[6] *Research Center for Neuromorphic AI hardware, Kyushu Institute of Technology, Kitakyushu 808-0196, Japan*

[#]Present address: *Birck Nanotechnology Center, School of Electrical and Computer Engineering, Purdue University, West Lafayette, Indiana 47907, USA.*

[*]Correspondence should be addressed to Y.F. (fukuma@cse.kyutech.ac.jp)

# Contents:

## S1. Estimation of $H_k$ and $H_p$ in CoFeB/MgO junction

In order to estimate the values of in-plane and perpendicular magnetic anisotropy fields i.e. $H_k$ and $H_p$, respectively, in our samples, we measured the in-plane azimuthal angle $\phi$ dependence of rectified voltage $V_{\text{ISHE}}$ in the samples used in the main text at an input power of $P_{\text{in}} = 0.001$ W. Low input power is essential to eliminate non-linear effects that can reduce accuracy of estimating anisotropy fields. The spectra were fitted using the Lorentzian function, as shown in Fig. S1a. From the fitting, we are able to estimate the values of resonance field $H_{res}^1$ as function of $\phi$ as shown in Fig. S1b. The data is then fitted using the following equation [29]:

$$H_{res}^1 = -H_k + \frac{3}{2} H_k \sin^2 \varphi - \frac{(M_s - H_p)}{2}$$

$$+ \frac{1}{2}\left[H_k^2 \sin^4 \varphi + (M_s - H_p)^2 + 2(M_s - H_p)H_k \sin^2 \varphi + 4\left(\frac{f}{\mu_0 \gamma}\right)^2\right]^{\frac{1}{2}}, (S1)$$

where, $M_s$ is the saturation magnetization measured to be 1.5 T using superconducting quantum interference device (SQUID), $\gamma$ is the gyromagnetic ratio whose value is estimated to be 0.0298 GHz.mT$^{-1}$ [29] and $\mu_0$ the permeability of free space. From this fitting, we can estimate $\mu_o H_p$ and $\mu_o H_k$ are 1405 mT and 2.5±0.2 mT in the samples.

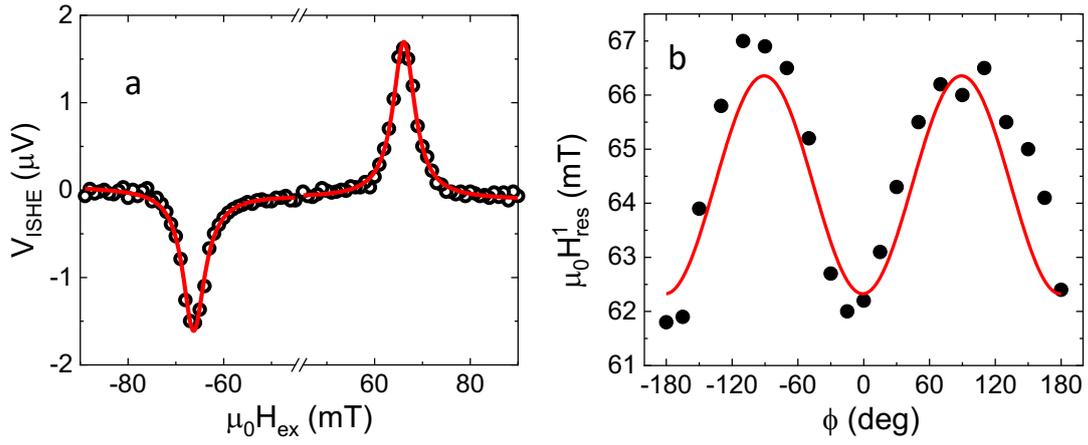

**Figure S1: a,** Inverse spin Hall effect voltage $V_{ISHE}$ spectrum at f = 3 GHz for an in-plane $H_{ex}$ along $\phi$ = 40 degree. Lines are fit to Lorentizian function to obtain resonance fields $H^1_{res}$. **b,** Azimuthal angle $\phi$ dependence of $H^1_{res}$ for f = 3 GHz. Sinusoidal dependence shows the presence of an in-plane anisotropy in addition to the perpendicular magnetic anisotropy. Line is fit using Eq. (S1).

## S2. Inverse spin Hall effect as origin of rectified voltage in multilayer stacks

We performed an in-plane angle dependent measurement of the rectified voltage to determine the origin of the rectified voltage in the Ta/Ru/Ta/CoFeB/MgO multilayer stacks. The samples used in this measurement are similar with our previous study [29]. Figure S2a shows the schematic of the setup used for detecting the rectified voltage in this process. FMR is excited by an z-axial microwave field $h_{\text{rf},z}$ that is generated due to a charge current flowing in a nearby coplanar waveguide (CPW). According to Harder *et al.* [46], in this setup, anisotropic magnetoresistance (AMR) and anomalous Hall effect (AHE) of the ferromagnet shows Lorentzian and dispersive lineshape, respectively, and spin pumping-inverse spin Hall effect (SP-ISHE) show Lorentzian lineshape. However, the AMR and ISHE spectra show a different symmetry as function of in-plane angle $\phi$. While the rectified voltage originating from AMR shows a $\sin 2\phi$ dependence, the SP-ISHE signal shows a $\cos\phi$ dependence. Figure S2b shows the rectified voltage in the sample. We can see a clear Lorentzian component in the spectrum. Figure S2c shows the angular dependent values for the Lorentzian component of the rectified voltage, which indicates a $\cos\phi$ dependence. This allows us to eliminate the possibility of AMR and AHE in our multilayer stacks and confirm that the rectified voltage originates from ISHE.

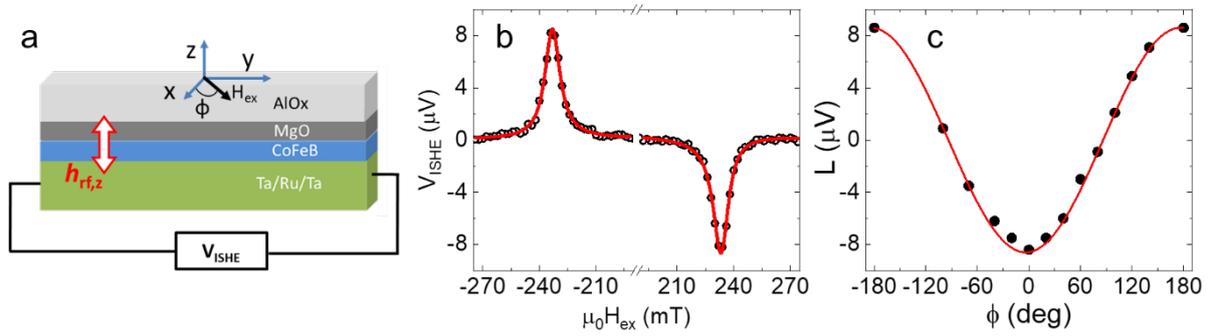

**Figure S2 a,** Experimental setup used for detecting rectified voltage across heavy metal underlayer. A z-axial microwave field $h_{rf,z}$ is generated by a coplanar waveguide near the multilayer strip and a magnetic field $H_{ex}$ is applied in the in-plane direction. **b,** Rectified voltage spectra (circles) obtained at a microwave frequency of 8 GHz. Lines are fits to the experimental data using the Lorentzian function. **c,** Amplitude of the Lorentzian $L$ component (circles) as a function of in-plane azimuthal angle $\phi$ with fit (line) to the experimental data using $\cos\phi$.

## S3. Estimation of threshold power for parametric excitation process

In order to estimate the threshold power for parametric pumping via electric-field torque, the $V_\text{ISHE}$ spectra are measured as a function of $P_\text{in}$ with an increment of 0.5 dBm as shown in Fig. S3. In this setup, up to an input power of 8.5 dBm (i.e. 7.1 mW), we can observe only one peak for the FMR mode, as can be seen in Fig. S3a. The background signal at this power in the spectra is around 6 μV and increases with input power to 15 μV at 11.5 dBm (i.e. 14.1 mW). The spectra presented in Fig. S3 have been vertically translated by subtracting such a background signal. At $P_\text{in}$ = 9.0 dBm (i.e. 7.9 mW), in addition to the FMR peak at $\mu_0 H_\text{ex}$ ~ 36 mT, we observe a new peak of the $V_\text{ISHE}$ spectrum around $\mu_0 H_\text{ex}$ ~ 12 mT over the background as shown in Fig. S3b. This value of $P_\text{in}$ is taken as the threshold power for $\mu_0 H_\text{ex}$ ~ 12 mT. The parametric peak broadens with increasing $P_\text{in}$, thereby allowing us to obtain the thresholds for other values of $H_\text{ex}$. As $P_\text{in}$ is increased to 11.5 dBm (i.e. 14.1 mW), we can obtain peaks for 0 mT < $\mu_0 H_\text{ex}$ < 15 mT in the spectra as shown in Fig. S3f. Using this process, the values of the threshold power are plotted as a function of $H_\text{ex}$ in Fig. 3a.

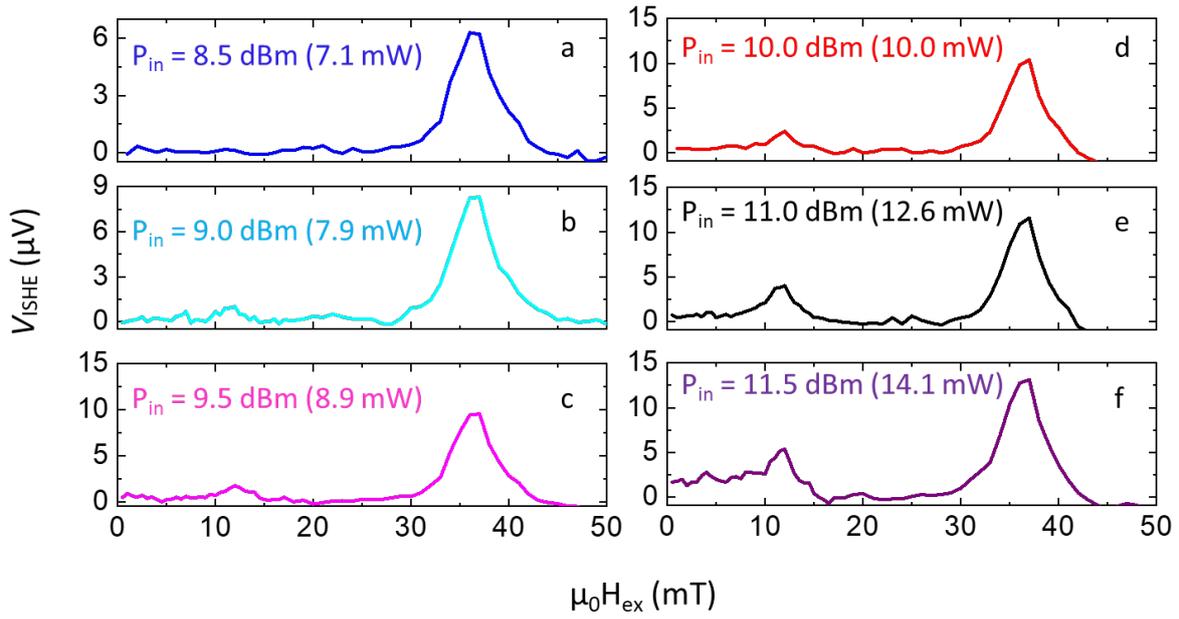

**Figure S3** Typical $V_{ISHE}$ spectra at $f = 2$ GHz as a function of input power $P_{in}$ ranging from 8.5 dBm to 11.5 dBm in the samples. Magnetization is oriented along the x-axis by applying $H_{ex}$ along $\theta = \phi = 0$ degree.

## S4. Inverse spin Hall voltage generated by electric-field torque and Oersted-field torque

The torque on magnetization in ferromagnets can be estimated using the relation $\tau \propto \hat{m} \times H_{eff}$, where $\hat{m}$ is the unit magnetization vector with components ($\cos\theta\cos\phi$, $\cos\theta\sin\phi$, $\sin\theta$) and $H_{eff}$ is the effective magnetic field given by the differential of magnetic free energy $F$ as follows:

$$H_{eff} = \frac{1}{\mu_o M} \frac{\partial F}{\partial \hat{m}}. \quad (S2)$$

The generalized expression of $F$ can be written as the sum of anisotropy, demagnetizing and Zeeman energies. For the sake of simplicity, we consider only the terms that have a time dependent characteristic as follows:

$$F = \frac{\mu_o M}{2}\left[-\frac{\partial H_p}{\partial V}V_{rf}m_z^2 - \frac{\partial H_k}{\partial V}V_{rf}m_x^2 + h_{rf,x}m_x + h_{rf,y}m_y + h_{rf,z}m_z\right], \quad (S3)$$

where $\frac{\partial H_p}{\partial V}V_{rf}$ and $\frac{\partial H_k}{\partial V}V_{rf}$ are the voltage modulated parts of $H_p$ and $H_k$ respectively, $h_{rf,x}$, $h_{rf,y}$ and $h_{rf,z}$ are the x-, y- and z-axial microwave fields and $m_x$, $m_y$ and $m_z$ are the x-, y- and z-components of $\hat{m}$. Due to the electric-field effect on the anisotropy fields, we can assume the anisotropy fields as the sum of $H_{ani} = (H_{ani})_{V=0} + \frac{\partial H_{ani}}{\partial V}V_{rf}$, where the first term on the right is the anisotropy in the unbiased condition i.e V = 0, while the second term accounts for the voltage modulated increase or decrease of $H_p$ or $H_k$ due to a microwave voltage of amplitude $V_{rf}$. Using Eq. (S2) and the VCMA terms of Eq. (S3), we can estimate the torque arising from VCMA of both $H_p$ and $H_k$ as:

$$\tau_{VCMA} \propto \begin{bmatrix} -\frac{\partial H_p}{\partial V} V_{rf} \sin\varphi \sin\theta \cos\theta \\ \left(\frac{\partial H_p}{\partial V} - \frac{\partial H_k}{\partial V}\right) V_{rf} \cos\varphi \sin\theta \cos\theta \\ \frac{\partial H_k}{\partial V} V_{rf} \sin\varphi \cos\varphi \cos^2\theta \end{bmatrix}. (S4)$$

It can be seen that due to the voltage modulation of $H_k$, we can have a non-zero torque on magnetization even when it is oriented in the film plane ($\theta = 0$ degree). Such a torque is not possible from the voltage modulation of $H_p$.

From the theory of spin pumping, DC component of inverse spin Hall effect voltage $V_{ISHE}$ in our device geometries is given by $V_{ISHE} = A\theta_c^2 \cos\phi \cos\theta$. Here, $\theta_c$ is the precession cone angle and $A = -\frac{\theta_{SH} e f \lambda_{sd} L_{NM} g^{\uparrow\downarrow}}{\sigma_{NM} t_{NM}} \tanh\left(\frac{t_{NM}}{2\lambda_{sd}}\right)$, $e$ being the charge of an electron and $g^{\uparrow\downarrow}$ is the spin mixing conductance of CoFeB/Ta interface. Precession cone angles are strongly dependent on the torque on magnetization. Assuming a linear relationship between the two in the linear excitation regime, i.e. $\theta_c = \eta\tau$, we can obtain the following expression for the angular dependent behavior if $V_{ISHE}$ originates from electric-field induced FMR:

$$V_{ISHE,VCMA} = AV_{rf}^2 \left(\left(\frac{\partial H_p}{\partial V}\sin\varphi\sin\theta\right)^2 + \left(\left(\frac{\partial H_p}{\partial V} - \frac{\partial H_k}{\partial V}\right)\cos\varphi\sin\theta\right)^2\right.$$
$$\left. + \left(\frac{\partial H_k}{\partial V}\sin\varphi\cos\varphi\cos\theta\right)^2\right)\cos\varphi\cos^3\theta. (S5)$$

Following a similar process, we can obtain the corresponding relations for $V_{ISHE}$ arising out $h_{rf,x}$, $h_{rf,y}$ and $h_{rf,z}$ as follows:

$$V_{ISHE,hx} = A\frac{h_{rf,x}^2}{4}(\sin^2\theta + \cos^2\theta \sin^2\varphi)\cos\varphi\cos\theta, (S6)$$

$$V_{ISHE,hy} = A\frac{h_{rf,y}^2}{4}(\sin^2\theta + \cos^2\theta \cos^2\varphi)\cos\varphi\cos\theta, (S7)$$

$$V_{ISHE,hz} = A\frac{h_{rf,z}^2}{4}\cos\varphi\cos^3\theta. (S8)$$

## S5. Oersted field distribution in Ta/CoFeB/MgO/Al$_2$O$_3$ multilayer strips.

In order to understand the Oersted field caused by an induced current when a microwave voltage is applied to the sample, we calculated the magnetic field distribution by using the finite element method (FEM) with COMSOL Multiphysics simulator. For simplicity, the Ta(5nm)/Ru(10nm)/Ta(5nm) buffer layer has been taken in the form of a single heavy metal layer of 20 nm Ta. As shown in Fig. S4a, the calculated sample size has been assumed to be 0.1 × 1 μm$^2$. The mesh grids were taken as triangular elements with a dimension of about 10 nm on a side. The input microwave signal of $f$ = 3 GHz is applied at a power of 10 μW on the upper Al$_2$O$_3$ surface while the bottom Ta surface is grounded. The applied power is 10$^4$ times smaller than the actual input power because the calculated sample size is 10$^4$ times smaller than the actual sample. Supplementary Figure 4b shows the color coded electric-field intensity as a function of the thickness. Additionally, the magnetic field distribution is observed to be along the longer axis of the sample as shown by arrows in the same diagram. We confirm this behavior by checking the Oersted field distribution in the samples with different length $l$ and width $w$ ratios as shown in Fig. S4c to S4e. Contributions from the z-axial Oersted field $h_z$ can be completely neglected in our samples. Moreover, it can be seen that as the $l$:$w$ increases, the component of Oersted field $h$ along the longer axis increases i.e. $h_y \gg h_x$.

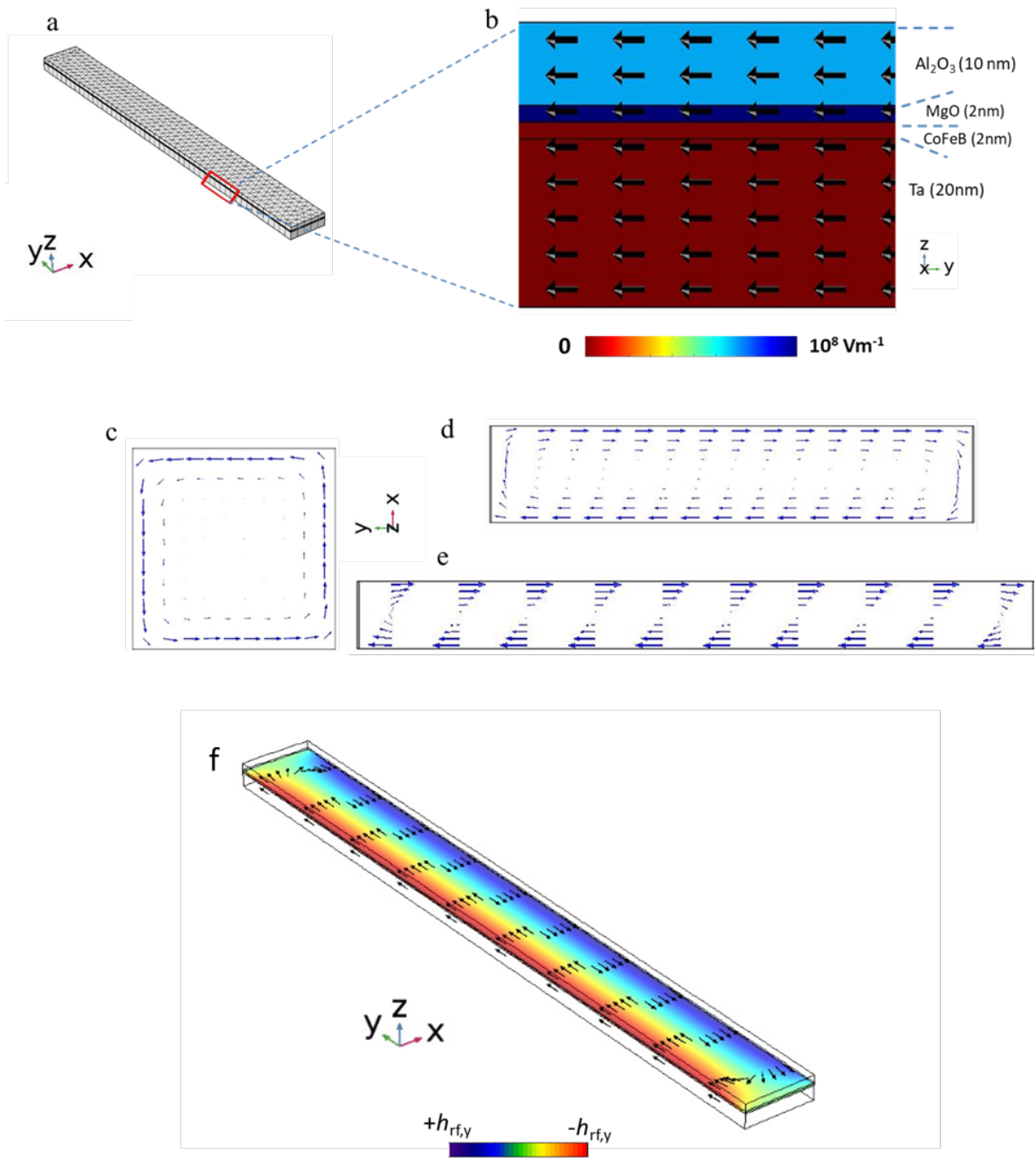

**Figure S4: a,** 20 nm Ta/ 2 nm CoFeB/ 2 nm MgO/ 10 nm $Al_2O_3$ multilayer structure with size $1 \times 0.1$ μm$^2$ discretized into triangular mesh elements of size around 10 nm. **b,** Electric-field distribution in the multilayer structure when a power of 10 μW is applied on the top surface is color coded as a function of the thickness while the corresponding magnetic field is shown as arrows. Normalized Oersted field distribution (arrows) is shown for samples with $l{:}w =$ **c,** 1:1 **d,** 1:5 and **e,** 1:10. **f,** The microwave field direction (arrows) and amplitude of $h_{rf,y}$ (color scale) in case of a sample with same $l{:}w$ as the experimental device i.e. 1:12.

## S6. Estimation of contributions of electric-field torque and Oersted-field torque from impedance measurement.

The reflection coefficient $S_{11}$ of the samples was measured using a vector network analyzer. Typical $S_{11}$ spectra in this study are shown in Fig. S5a. In order to estimate the impedance of the CoFeB/MgO junction, we assume an equivalent circuit which is the same model for electric-field induced FMR study reported by Nozaki *et al.* [23], as shown in Fig. S5b. The $S_{11}$ spectra are fitted using the expression $S_{11} = \dfrac{Z - 50\ \Omega}{Z + 50\ \Omega}$. Here, 50 Ω is the characteristic impedance of the measurement set up, while $Z$ is the net impedance obtained from the sample with inductive component $L$, resistance $R_0$ and capacitance $C$ of the junction, resistive loss between the top and bottom electrodes $r$. From the experimental data we can see that the Real[$S_{11}$] → 1 when frequency → 0. This indicates a very high value of $R_0$ because of the 2-nm thick MgO and 10-nm thick $Al_2O_3$ layers in the sample. The $S_{11}$ spectrum in Fig. S5a is fitted using $R_0 = 10^5\ \Omega$, $r = 30\ \Omega$, $C = 3.5$ pF and $L = 70$ pH.

Using the $S_{11}$ spectra, we can estimate the contributions of the electric-field torque and the Oersted-field torque as a function of $f$ for the sample. In the case of the electric-field torque, the voltage applied to the sample is expressed as $V \propto (1 + S_{11})V_{input} = \zeta_V V_{input}$. In the case of the Oersted-field torque, the current is proportional to the square root of the power inserted into the sample, as $I \propto \sqrt{P} = \sqrt{(1 - |S_{11}|^2)P_{in}} = \zeta_I \sqrt{P_{in}}$. The efficiency of the electric-field torque decreases with increasing $f$, on the other hand the efficiency of the Oersted-field torque increases with increasing $f$.

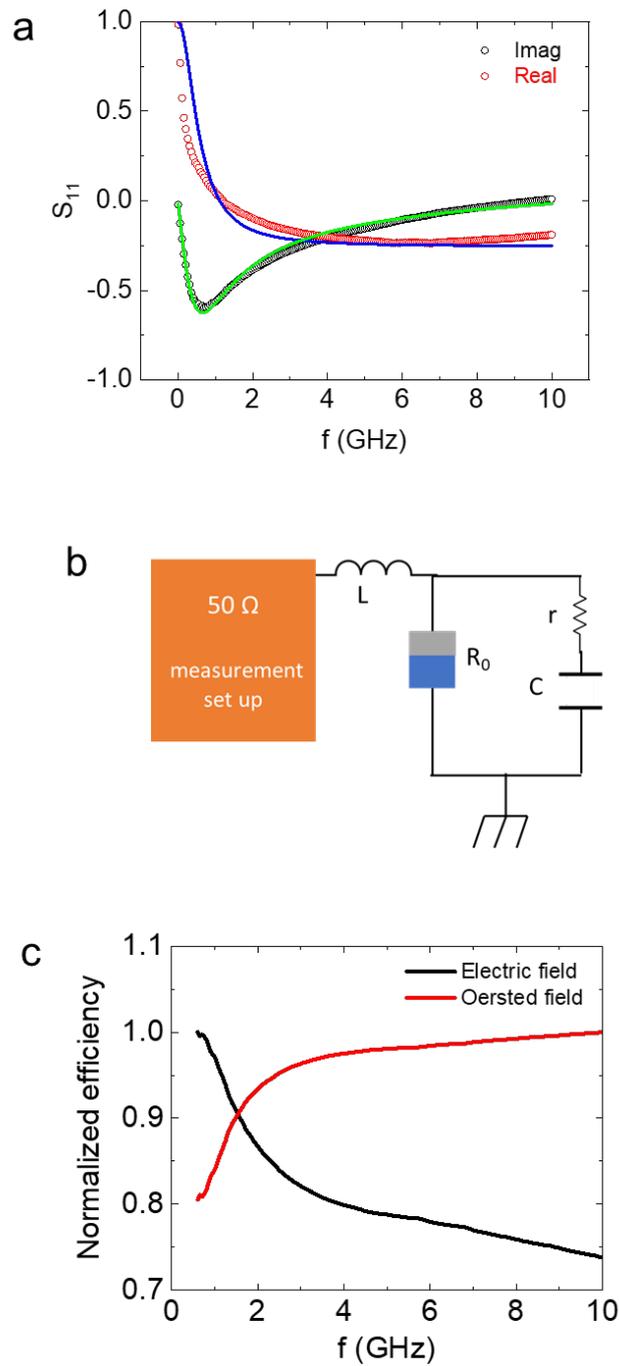

**Figure S5: a,** Typical $S_{11}$ spectra of the sample in this study. Lines are fits to $S_{11} = \dfrac{Z - 50\ \Omega}{Z + 50\ \Omega}$. **b,** Schematic of the equivalent circuit of the sample in this measurement. **c,** Normalized efficiencies calculated for the electric-field torque and the Oersted-field toque as a function of frequency of $P_{in}$.

## S7. Angular dependence of rectified voltage in higher frequencies.

We performed frequency dependent angular ISHE measurements to further establish that our samples show electric-field excitation of magnetization dynamics. Figure S6(a), shows the $V_{\text{ISHE}}$ at frequencies 2 and 4 GHz respectively at $\theta = 0$ degree, $\phi = 40$ degree at $P_{\text{in}} = 0.001$ W. The angular dependent data for $f = 4$ GHz is presented in Fig. S6(b). Upon fitting the data to Eq.(4) of the main text, we obtained values $A\left(\dfrac{\partial H_k}{\partial V}V_{rf}\right)^2 = 0.90$ and $Ah_{rf,y}^2 = 0.65$. Note that from the data in Fig. 5(a) of main text, we obtained $A\left(\dfrac{\partial H_k}{\partial V}V_{rf}\right)^2 = 2.87$ and $Ah_{rf,y}^2 = 0.88$ for $f = 2$ GHz. It shows that as frequency increases, the VCMA contribution is reduced, while the Oersted-field contribution increases in our devices. This further proves the presence of VCMA of $H_k$ in our devices, as can be understood from the previous section S6, which shows that their torques have a decreasing and increasing frequency dependence, respectively.

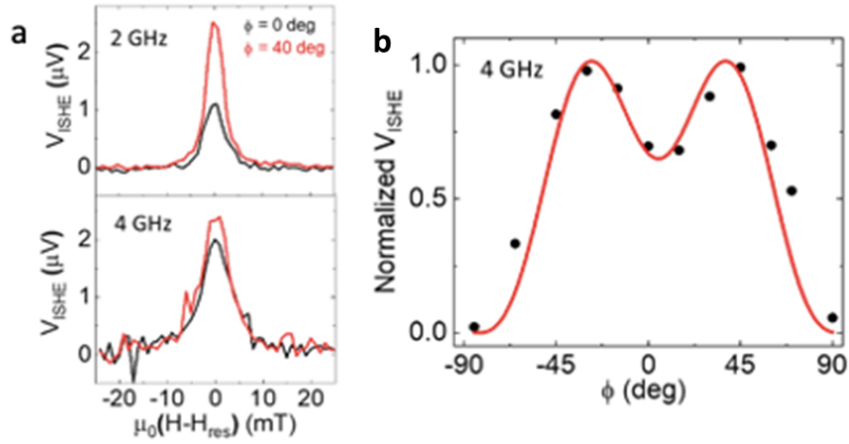

**Figure S6: a,** Rectified voltages for 2 and 4 GHz at $\phi = 0$ and 40 degrees. **b,** Normalized $\phi$ dependence of $V_{\text{ISHE}}$ for 4 GHz. Line is fit to Eq.(4) of main text using values $A\left(\dfrac{\partial H_k}{\partial V}V_{rf}\right)^2 = 0.90$ and $Ah_{rf,y}^2 = 0.65$.

## S8. Power dependence of rectified voltage in control sample.

We measured the input power dependence of the $V_{ISHE}$ spectra for control samples (see Fig. S2a for the sample structure) via the Oersted-field torque excitation. Figure S6a shows a typical power dependent $V_{ISHE}$ spectra at $f=3$ GHz and $\phi = \theta = 0$ degree. In this configuration, $h_{rf,z}$ excites FMR mode with a maximum torque, as can be seen in Eq. (S8). Figure S6b shows the $V_{ISHE}$ as a function of input power $P_{in}$ which confirms that the sample enters non-linear excitation regime at $P_{in} > 0.07$ W. However, no peak in Fig. S6a can be observed for the parametric excitation process. This indicates the perpendicular pumping shows a high threshold power of the parametric exciation of the Ta/CoFeB/MgO structure.

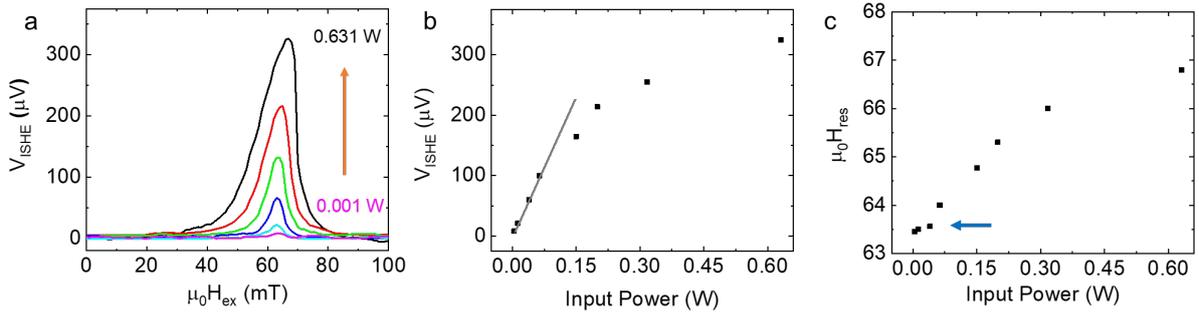

**Figure S7: a,** Typical $V_{ISHE}$ spectra as a function of input power $P_{in}$ ranging from 0.001 W to 0.631 W at $f = 2$ GHz. **b,** $V_{ISHE}$ amplitudes and **c,** Resonance field $H_{res}$ as a function of the input power. Linear fit (grey line) in **b** implies the transition from linear to non-linear regime.

**References:**

[46] M. Harder, Y. Gui, and C-M. Hu, *Phys. Rep.* **661**, 1-59 (2016).